\documentclass{article}
\usepackage{amsmath}
\usepackage[latin1]{inputenc}
\usepackage[super]{cite}
\usepackage{graphicx} 
\usepackage{perpage}
\MakePerPage{footnote}

\begin{document}
\markboth{ANTONIO DI LORENZO}
{ARE QUANTUM CORRELATIONS GENUINELY QUANTUM?}

%
%
\title{ARE QUANTUM CORRELATIONS GENUINELY QUANTUM?}
\author{Antonio Di Lorenzo\\
Instituto de F\'{\i}sica, Universidade Federal de Uberl\^{a}ndia, \\ Av. Jo\~{a}o Naves de \'{A}vila 2121, 
Uberl\^{a}ndia, Minas Gerais, 38400-902,  Brazil\\ dilorenzo@infis.ufu.br}
\date{}
\maketitle
\begin{abstract}
It is shown that the probabilities for the spin singlet can be reproduced through classical resources, with no communication between the distant parties, by using merely shared (pseudo-)randomness. 
If the parties are conscious beings aware of both the hidden-variables and the random mechanism, then one has a conspiracy. 
If the parties are aware of only the random variables, they may be induced to believe that they are able to send instantaneous information to one another. 
It is also possible to reproduce the correlations at the price of reducing the detection efficiency. 
It is further demonstrated that the same probability decomposition could be realized through  
action-at-a-distance, provided it existed. 
\end{abstract}

\section{Introduction and Outlook}
Entangled states, first considered in Ref.~\citeonline{Einstein1935}, show one of the most challenging feature 
of quantum mechanics: unmeasured observables have no value. This fact manifests in peculiar probability 
distributions that do not admit embedding in a master, positive-definite probability that assigns simultaneous values 
to incompatible observables. While this statement applies as well to single particle states, that do not have in general 
a positive Wigner function, this feature becomes particularly interesting when components of an entangled state 
are separated in space. Then it is possible to establish the incompatibility between quantum mechanics and not one, 
but whole families of possible more fundamental theories, called hidden variable theories, that satisfy some general hypotheses. 
This remarkable feat was achieved by Bell,\cite{Bell1964} who showed that a family of deterministic theories 
predicts that spin correlators satisfy an inequality, while the corresponding correlators predicted by quantum mechanical do not. 
Later the result was extended to a family of stochastic theories.\cite{Clauser1974}
Experimental evidence excludes these theories\cite{Freedman1972,Aspect1982a,Aspect1982b,Tapster1994,Weihs1998,Tittel1998,Rowe2001}, 
so that, if a completion of quantum mechanics is to be found, one must look at theories that violate at least one 
of the hypotheses leading to Bell inequality. 
While locality, intended here as the impossibility of action-at-a-distance, is one of these hypotheses, it is not the only 
one, despite a widespread belief that the present manuscript will prove unfounded. 
We are left then with the challenge of finding theories that reproduce the quantum mechanical correlations. 
In this paper, we concentrate on theories that violate the hypothesis of Uncorrelated Choice, 
which establishes the absence of correlations between the settings of the measuring instrument and the hidden variables. 
This violation admits several interpretations: a conspiracy between the measuring parties, a limitation of their free will, 
a non-local influence of the settings on the hidden variable, and also a conspiracy between the particles. 
We discuss also how to verify which interpretation is the correct one. 
In particular, we propose an (ideal) experiment allowing to decide whether we are in presence of non-locality or 
of limited free will, assuming that the hidden variables are experimentally accessible (if they were not, why bother?).
This experiment, however, cannot rule out a superdeterminist theory, where a cosmic choreographer has decided from the beginning of times 
not only how experimentalists will choose the settings of their apparatuses (or which apparently-to-them random devices they select), 
but also all arbitrary conventions assigning meaning to, e.g., spin up and spin down, etc.\footnote{Superdeterminism is unfalsifiable, but I assign it a very small probability of 
being true. It is akin to solipsism or intelligent design, in that it can explain anything but predict nothing. With Schopenhauer (who was talking about solipsism), 
I would say that superdeterminism should be treated as a \lq\lq{}small frontier fortress. Admittedly the fortress is impregnable, but the garrison can never sally forth from it, and therefore we can pass it by and leave it in our rear without danger.\rq\rq{}\cite{Schopenhauer}} 
Finally, given that the model we propose can be realized through local classical resources, and that it reproduces the quantum mechanical 
probabilities, it is to be concluded that quantum correlations are not exclusively quantum.

After reviewing the typical setup in Section \ref{sec:setup}, in Section \ref{sec:bellhyp} we  
discuss the hypotheses at the basis of the derivation of Bell and Leggett-type inequalities, 
dispelling some misconceptions about their physical meaning. 
Section \ref{sec:bellboole} presents a simpler derivation of Bell inequality, based on a procedure pioneered by Boole 
relying only on the hypothesis of counter-factual definiteness. 
This derivation actually predates Bell\rq{}s, but it was discovered 
in a mathematical context, 
where it was not realized the depth of its implications in physics, until it was rediscovered several years later. 
Furthermore, we discuss how this alternative derivation has a great relevance in refuting the Einstein-Podolski-Rosen argument, 
as it gives quantitative support to the criticisms of Bohr and others. 
Section \ref{sec:tb} analyzes the Toner and Bacon model,\cite{Toner2003} showing that it can be obtained as the limiting 
case of two different families of models, each violating a different Bell hypotheses, and also that it can be realized in a 
natural way through a limitation of the possible choices of polarization at one station.
Section \ref{sec:hall} presents a model by M.J.W. Hall,\cite{Hall2010} based on minimal correlations between the hidden parameters and the 
choice of polarization. 
Section \ref{sec:model} introduces a model\cite{DiLorenzo2012b} violating only one 
of the Bell hypotheses and retaining Malus\rq{}s law. 
It is also discussed how the model proposed can be explained in various ways --- which include 
\lq\lq{}conspiracy\rq\rq{}, limitation of \lq\lq{}free will\rq\rq{}, exploitation of the detection loophole, or action-at-a-distance,--- depending on what additional hypotheses are made 
on the physical nature of the hidden variables.
The two models of Section \ref{sec:hall} and Section \ref{sec:model} are combined in Section \ref{sec:cirel}, yielding a model that can violate 
Cirel\rq{}son bound and reach the maximum violation of Bell inequality. 
Section \ref{sec:inter} discusses the relation between non-locality and \lq\lq{}slave will\rq\rq{}, and shows that it is possible to distinguish between the 
two, provided the hidden parameters can be measured, barring superdeterminism. 
Appendix \ref{app:locdef} discusses competing definitions of locality, and shows with one example that one of them, which is used routinely in discussions of Bell inequality and 
that Bell argued coincided with \lq\lq{}local causality\rq\rq{}, conflicts with the other ones. 
Appendix \ref{app:real} shows how it is possible to implement the models of Sections \ref{sec:hall} and \ref{sec:model} through local classical resources, with no 
communication nor shared randomness between the distant parties, by only having shared randomness between each party and the entangler. 

\section{The system}\label{sec:setup}
\begin{figure}
\centering
\includegraphics[width=4in]{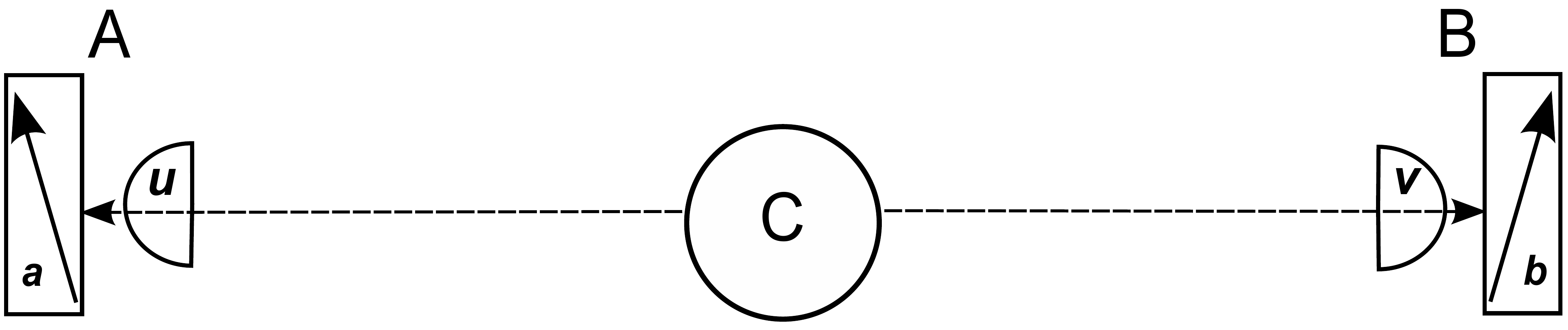}
\caption{\label{fig:eprbsetup} Scheme of the setup considered. Regions A and B are spacelike separated. The semicircles represent hypothetical detectors for the hidden variables as discussed in the Sec. \ref{sec:inter}.}
\end{figure}
The setup considered (see Fig. \ref{fig:eprbsetup}) consists of two particles produced in a region $C$ of the 
spacetime, each travelling to a different detection region, $A$ and $B$. The measurement of each particle does 
not need to happen at the same time, but the two detection events are assumed to have spacelike separation, 
so that a reference frame exists in which the measurements are simultaneous. 
The outcomes of the two measurements are two-valued, and will be denoted by 
$e_A=\sigma=\pm 1$ and $e_B=\tau=\pm 1$. 
The $A$ ($B$) detector is characterized by a unit vector $\mathbf{n}_A=\mathbf{a}$ ($\mathbf{n}_B=\mathbf{b}$), corresponding to the 
orientation of a spin measuring device in Quantum Mechanics. 
The quantity of interest is the joint probability $P(\sigma,\tau|\Psi,\Sigma)$
with $\Psi$ describing the preparation of a singlet state, and $\Sigma=\{\mathbf{a},\mathbf{b}\}$ 
specifying the observables being measured. 
Since $\Psi$ appears as a prior in all the probabilities considered, it is omitted for brevity. 
We shall write for brevity $P(\sigma,\tau|\Psi,\Sigma)=P_{\sigma,\tau}(\Sigma)$. 
A hidden variable theory consists in assuming the existence of additional parameters, $\lambda$, providing a finer description 
of the state of the system, so that, according to Bayes\rq{}s rule 
\begin{equation}
P_{\sigma,\tau}(\Sigma)=\int d\mu(\lambda|\Sigma) P_{\sigma,\tau}(\lambda,\Sigma) ,
\end{equation}
where $\mu(\lambda|\Sigma)=\mu_\lambda(\Sigma)$ is the probability distribution for the additional parameters under 
the experimental conditions $\Sigma=\{\mathbf{a},\mathbf{b}\}$. 
The joint probability $P_{\sigma,\tau}(\lambda,\Sigma)$ can be further decomposed 
in either of two ways: 
\begin{equation}
P_{\sigma,\tau}(\lambda,\mathbf{a},\mathbf{b})=
M_{\sigma}^A(\lambda,\mathbf{a},\mathbf{b}) Q_{\tau}^B(\lambda,\mathbf{b},\mathbf{a},\sigma) 
=M_{\tau}^B(\lambda,\mathbf{b},\mathbf{a}) Q_{\sigma}^A(\lambda,\mathbf{a},\mathbf{b},\tau) , 
\end{equation} 
with the marginal probabilities 
\begin{subequations}
\begin{align}
M_\sigma^A(\lambda,\mathbf{a},\mathbf{b})\equiv&  
\sum_{\tau} P_{\sigma,\tau}(\lambda,\mathbf{a},\mathbf{b}), \\
M_\tau^B(\lambda,\mathbf{b},\mathbf{a})\equiv&  
\sum_{\sigma} P_{\sigma,\tau}(\lambda,\mathbf{a},\mathbf{b}),
\end{align}
\end{subequations}
and the conditional probabilities  
\begin{subequations}\label{eq:bayes}
\begin{align}
Q^A_\sigma(\lambda,\mathbf{a},\mathbf{b},\tau)\equiv& 
\frac{P_{\sigma,\tau}(\lambda,\mathbf{a},\mathbf{b})}{M^B_{\tau}(\lambda,\mathbf{b},\mathbf{a})} \ , \\
Q^B_\tau(\lambda,\mathbf{b},\mathbf{a},\sigma)\equiv& 
\frac{P_{\sigma,\tau}(\lambda,\mathbf{a},\mathbf{b})}{M^A_{\sigma}(\lambda,\mathbf{a},\mathbf{b})} \ . 
\end{align}
\end{subequations}
For models respecting the symmetry between $A$ and $B$, if $B$ measured along $\mathbf{a}$ and 
$A$ along $\mathbf{b}$ the new probabilities should coincide with the former ones, i.e., 
$M_\sigma^A(\lambda,\mathbf{a},\mathbf{b})=M_\sigma^B(\lambda,\mathbf{a},\mathbf{b})$ and 
$Q^A_\sigma(\lambda,\mathbf{a},\mathbf{b},\tau)=Q^B_\sigma(\lambda,\mathbf{a},\mathbf{b},\tau)$, 
so that the upper index can be dropped since the functions coincide. 
There are models, however, where this symmetry is lost at the hidden variables level, see, e.g., the Toner and Bacon model discussed 
below. 
In the following, we shall write the hidden variables as $\lambda=\lambda_A\cup\lambda_B\cup\lambda_G$, 
where $\lambda_j$ ($j\in\{A,B\})$ refers to the local parameters, i.e. 
to the variables that can be associated, at the time of detection, to the regions $A$ and $B$ (they can be parameters locally 
attached to the particles and parameters describing the detection regions) and 
$\lambda_G$ refers to the global parameters, i.e., parameters that it is not possible to pin down to a region of 
space-time, as the wavefunction of an entangled system.
Bell\cite{Bell1976a} 
used the term \lq\lq{}beables\rq\rq{} to refer to local parameters, and assumed that there are no global parameters, 
presumably motivated by classical mechanics, where the local parameters are dynamical quantities describing 
physical systems, and no global parameters are needed if all the local ones are known with infinite precision.
\section{Hypotheses at the basis of Bell and Leggett inequalities}\label{sec:bellhyp}
\subsection{Bell inequality}
Events corresponding to the detection of one particle, e.g., the determination of the spin projection of an electron 
along a given direction, can be reproduced easily by a hidden-variable model satisfying 
some reasonable requirements\cite{Bell1964}. However, when one considers a system composed of two entangled 
particles, reproducing the predictions of quantum mechanics becomes a non-trivial matter. 
Refs.~\citeonline{Bell1964} and \citeonline{Clauser1969} 
demonstrated that a family of deterministic and local models is incompatible with quantum mechanics, 
and the experiments\cite{Freedman1972,Aspect1982a,Aspect1982b,Tapster1994,Weihs1998,Tittel1998,Rowe2001}  
have corroborated this statement by reproducing the quantum mechanical predictions with remarkable accuracy. 
Refs.~\citeonline{Bell1964} and \citeonline{Clauser1969} discriminated between quantum mechanics 
and the said special class of deterministic local 
hidden-variable theories through inequalities, known as Bell and CHSH inequalities. 
The CHSH inequality, in particular, proves to be easier to check against experimental data, and it is currently written 
in a different form than in the original paper\cite{Clauser1969} as 
\begin{equation}
|C(\mathbf{a},\mathbf{b})+C(\mathbf{a}\rq{},\mathbf{b})+C(\mathbf{a},\mathbf{b}\rq{})-C(\mathbf{a}\rq{},\mathbf{b}\rq{})|
\le 2,
\end{equation} 
with $C$ the correlator (assuming $\overline{\sigma}=\overline{\tau}=0$)
\begin{equation}
C(\mathbf{a},\mathbf{b})=\sum_{\sigma,\tau} \sigma\tau P_{\sigma,\tau}(\mathbf{a},\mathbf{b}).
\end{equation}
Under the hypothesis of determinism, the marginal probabilities are
$M^A_\sigma(\lambda,\mathbf{a},\mathbf{b})\in\{0,1\}$ and  
$M^B_\tau(\lambda,\mathbf{b},\mathbf{a})\in\{0,1\}$. 
Then there is a discontinuous function $S$ (respectively, $T$) such that 
$\sigma_0=S(\lambda,\mathbf{a},\mathbf{b})$ (respectively, $\tau_0=T(\lambda,\mathbf{b},\mathbf{a})$) is the only 
permitted value of $\sigma$ (resp., $\tau$). 
Hence, the conditional probabilities are
\begin{subequations}
\begin{align}
Q^A_\sigma(\lambda,\mathbf{a},\mathbf{b},\tau_0)=&Q^A_\sigma(\lambda,\mathbf{a},\mathbf{b},T(\lambda,\mathbf{b},\mathbf{a}))=M^A_\sigma(\lambda,\mathbf{a},\mathbf{b}) , \\
Q^B_\tau(\lambda,\mathbf{b},\mathbf{a},\sigma_0)=&Q^B_\tau(\lambda,\mathbf{b},\mathbf{a},S(\lambda,\mathbf{a},\mathbf{b}))=M^B_\tau(\lambda,\mathbf{b},\mathbf{a}) .
\end{align}
\end{subequations}
Since the marginals are 
\begin{subequations}\label{eq:margdef}
\begin{align}
M^A_\sigma(\lambda,\mathbf{a},\mathbf{b})=&\sum_{\tau} P_{\sigma,\tau}(\lambda,\mathbf{a},\mathbf{b}) 
= P_{\sigma,\tau_0}(\lambda,\mathbf{a},\mathbf{b})=M^B_{\tau_0}(\lambda,\mathbf{b},\mathbf{a}) 
Q^A_\sigma(\lambda,\mathbf{a},\mathbf{b},\tau_0)
\\
M^B_\tau(\lambda,\mathbf{b},\mathbf{a})=&\sum_{\sigma} P_{\sigma,\tau}(\lambda,\mathbf{a},\mathbf{b}) 
= P_{\sigma_0,\tau}(\lambda,\mathbf{a},\mathbf{b})
=M^A_{\sigma_0}(\lambda,\mathbf{b},\mathbf{a}) 
Q^B_\tau(\lambda,\mathbf{b},\mathbf{a},\sigma_0),
\end{align}
\end{subequations}
we have that for deterministic theories 
\begin{subequations}\label{eq:oi}
\begin{align}
\label{eq:oia}
Q^A_\sigma(\lambda,\mathbf{a},\mathbf{b},\tau_0)=&M^A_\sigma(\lambda,\mathbf{a},\mathbf{b}) , \\
\label{eq:oib}
Q^B_\tau(\lambda,\mathbf{b},\mathbf{a},\sigma_0)=&M^B_\tau(\lambda,\mathbf{b},\mathbf{a}) .
\end{align}
\end{subequations}
For $\sigma\neq S(\lambda,\mathbf{a},\mathbf{b})$ or $\tau\neq T(\lambda,\mathbf{b},\mathbf{a})$, the conditional  
probabilities take the indeterminate form $\tfrac{0}{0}$. 
In any case, the joint probability, conditioned on $\lambda$, factorizes as 
\begin{equation}\label{eq:factor}
P_{\sigma,\tau}(\lambda,\mathbf{a},\mathbf{b})= M^A_\sigma(\lambda,\mathbf{a},\mathbf{b}) 
M^B_\tau(\lambda,\mathbf{b},\mathbf{a}) 
\end{equation}	
and the marginals are 
\begin{equation}\label{eq:deter}
M^A_\sigma(\lambda,\mathbf{a},\mathbf{b})=
\delta_{\sigma,S(\lambda,\mathbf{a},\mathbf{b})} \ , \ 
M^B_\tau(\lambda,\mathbf{b},\mathbf{a})=
\delta_{\tau,T(\lambda,\mathbf{b},\mathbf{a})}
\end{equation}

Bell\cite{Bell1964} then requires that the marginal probabilities obey the hypothesis of Setting Independence, 
\begin{subequations}
\label{eq:si}
\begin{align}
\label{eq:sia}
M^A_\sigma(\lambda,\mathbf{a},\mathbf{b})=&M^A_\sigma(\lambda,\mathbf{a})\\
\label{eq:sib}
M^B_\tau(\lambda,\mathbf{b},\mathbf{a})=&M^B_\tau(\lambda,\mathbf{b})\ .
\end{align}
\end{subequations}
For deterministic theories, Eqs.~\eqref{eq:si} are equivalent to 
 $S(\lambda,\mathbf{a},\mathbf{b})=S(\lambda,\mathbf{a})$, 
 $T(\lambda,\mathbf{b},\mathbf{a})=T(\lambda,\mathbf{b})$.
This hypothesis derives from the hypothesis of locality, meant as the impossibility of action-at-a-distance. 
Setting Independence is actually weaker than locality: let us imagine that the parameters $\lambda$ can be 
divided into localized parameters associated to the particle reaching $A$, $\lambda_A$, localized parameters 
associated to the particle reaching $B$, $\lambda_B$, and global parameters $\lambda_G$
that cannot be associated to a particle. Setting Independence allows the marginal probability at $A$ to depend 
on the remote local parameters $\lambda_B$, even though it forbids dependence on the remote parameter $\mathbf{b}$. 

Finally, it is required that 
\begin{equation}\label{eq:uc}
\mu(\lambda|\mathbf{a},\mathbf{b})=\mu(\lambda) .
\end{equation}
This hypothesis is sometimes called Measurement Independence, but we prefer to 
call it Uncorrelated Choice, meaning that the choice of the settings is uncorrelated to the value of $\lambda$. 
The original justification of Uncorrelated Choice relied on locality: 
if one imagines the $\lambda$ as a deterministic time-evolution of $\lambda_0$, parameters 
localized at the entangler when the particles are emitted, changing the settings 
can not influence $\lambda_0$ and hence it cannot influence $\lambda$.  
However, this picture was revealed\cite{Shimony1976a} to be inexact, since a correlation between two events 
coinciding in space-time 
does not imply necessarily a causal relation between them, and even when there is such a causal relation, 
which event is the cause and which the effect is merely metaphysical (at least for time-symmetric 
physical theories such as classical and quantum mechanics). 
Indeed, by having $\mu(\lambda|\mathbf{a},\mathbf{b})$ we can think that the $\lambda$ influence the choice 
of the settings, rather than \emph{vice versa}. 
Locality implies not Eq.~\eqref{eq:uc}, but only that 
\begin{equation}
\mu^A(\lambda_A|\lambda_G,\mathbf{a},\mathbf{b})=\frac{\int_{\Omega_B} d\lambda_B 
\mu(\lambda_A,\lambda_B,\lambda_G|\mathbf{a},\mathbf{b})}
{\int_{\Omega_A} d\lambda_A  \int_{\Omega_B} d\lambda_B 
\mu(\lambda_A,\lambda_B,\lambda_G|\mathbf{a},\mathbf{b})}=\mu^A(\lambda_A|\lambda_G,\mathbf{a}) 
\end{equation}
and 
\begin{equation}
\mu^B(\lambda_B|\lambda_G,\mathbf{b},\mathbf{a})=
\frac{
\int_{\Omega_A} d\lambda_A \mu(\lambda_A,\lambda_B,\lambda_G|\mathbf{a},\mathbf{b})}
{
\int_{\Omega_A} d\lambda_A\int_{\Omega_B} d\lambda_B  \mu(\lambda_A,\lambda_B,\lambda_G|\mathbf{a},\mathbf{b})}
=\mu^B(\lambda_B|\lambda_G,\mathbf{b}) ,
\end{equation}
where $\Omega_j$ is the domain of $\lambda_j$. 
Thus the hypothesis of Uncorrelated Choice is stronger than locality. 

In summary, the original derivation of Bell and CHSH inequalities relied on Determinism, which implies 
Eqs.~\eqref{eq:factor} and \eqref{eq:deter}, Setting Independence, which implies Eq.~\eqref{eq:si}, and on 
Uncorrelated Choice, which requires Eq.~\eqref{eq:uc}. 
Since Setting Independence and Uncorrelated Choice were believed to follow from locality, 
the inequalities, which are violated by quantum mechanics, 
were said to establish an incompatibility between local realistic theories and quantum mechanics. 
The term \lq\lq{}realism\rq\rq{} was used by Bell as synonymous with determinism. 
Later on, this statement was corrected to the incompatibility between quantum mechanics and all 
local deterministic theories allowing complete \lq\lq{}free will\rq\rq{}, i.e. satisfying Uncorrelated Choice. 

While Eq.~\eqref{eq:deter} was used in the original derivations of Refs.~\citeonline{Bell1964} and \citeonline{Clauser1969}, 
later it was realized\cite{Bell1971,Clauser1974} that it was unnecessary, so that the probabilities were allowed  
to take values in the whole range $[0,1]$. However, it was essential for deriving the inequalities, 
that Eq.~\eqref{eq:factor}, another consequence of determinism, still held. 
Confusingly, Eq.~\eqref{eq:factor} was instead considered a consequence of locality, perhaps because 
it was implicitly assumed that the hidden variables consisted only of local parameters. 
It was then claimed that an incompatibility existed between all stochastic, local hidden-variable models and quantum mechanics. 
Ref.~\citeonline{Jarrett1984} pointed out, however, that  
Eq.~\eqref{eq:oi} is not implied by locality, which, we remind the readers, means 
the impossibility of action-at-a-distance. 
Hence Eq.~\eqref{eq:oi} or equivalently Eq.~\eqref{eq:factor} 
has to be postulated for stochastic models.
Ref.~\citeonline{Jarrett1984} called such hypothesis \lq\lq{}completeness\rq\rq{}, while 
Ref.~\citeonline{Shimony1990} coined the name ``Outcome Independence''. 
On the other hand, Bell\cite{Bell1976a} postulated \lq\lq{}Outcome Independence\rq\rq{} as 
the definition of \lq\lq{}local causality\rq\rq{}. Appendix \ref{app:locdef} shows with an example that 
Bell\rq{}s definition is untenable, if one accepts that there are global parameters.  

We shall refer to Eq.~\eqref{eq:factor} as Reducibility of Correlations, since it means that the knowledge of $\lambda$ breaks 
down the correlations between $\sigma$ and $\tau$, so that correlations are reducible to ignorance of $\lambda$. 
Reducibility of Correlations implies that 
\begin{subequations}\label{eq:oi2}
\begin{align}
\label{eq:oi2a}
Q^A_\sigma(\lambda,\mathbf{a},\mathbf{b},\tau)=&\begin{cases}M^A_\sigma(\lambda,\mathbf{a},\mathbf{b}) \ , 
\forall \tau: M^B_\tau(\lambda,\mathbf{b},\mathbf{a}) \neq 0 \\
\tfrac{0}{0} \ , \forall \tau:
M^B_\tau(\lambda,\mathbf{b},\mathbf{a}) = 0
\end{cases} , \\
\label{eq:oi2b}
Q^B_\tau(\lambda,\mathbf{b},\mathbf{a},\sigma)=&\begin{cases}M^B_\tau(\lambda,\mathbf{b},\mathbf{a}) \ , 
\forall \sigma:
M^A_\sigma(\lambda,\mathbf{a},\mathbf{b}) \neq 0 \\
\tfrac{0}{0} \ , \forall \sigma:
M^A_\sigma(\lambda,\mathbf{a},\mathbf{b}) = 0
\end{cases} . 
\end{align}
\end{subequations}
These equations characterize Outcome Independence, and they in turn imply Reducibility of Correlations. 
Hence the two terms are interchangeable.\footnote{Strictly speaking, if one interprets the symbol $\tfrac{0}{0}$ as having a meaning of its own, then the term  
\lq\lq{}Outcome Independence\rq\rq{} does not describe properly Eqs.~\eqref{eq:oi2}. 
However, one could as well say that since the symbol $\tfrac{0}{0}$ is indeterminate, one may assign to it an 
arbitrary value, $\tfrac{0}{0}=M^A_\sigma(\lambda,\mathbf{a},\mathbf{b})$ and 
$\tfrac{0}{0}=M^B_\tau(\lambda,\mathbf{b},\mathbf{a})$.} 

In conclusion, Bell-type inequalities establish an incompatibility 
between quantum mechanics and a special class of stochastic models, the ones satisfying  
the hypotheses of Reducibility of Correlations, Setting Independence, and Uncorrelated Choice. 
While Ref.~\citeonline{Jarrett1984} and subsequent papers\cite{Howard1989,Jarrett1989,Redhead1989,Shimony1990}
highlighted the fact that Reducibility of Correlations is not 
a consequence of locality, these contributions were forgotten in the 90s, when the word \lq\lq{}locality\rq\rq{} 
in the context of Bell inequalities was redefined to mean compliance with the three hypotheses 
Reducibility of Correlations, Setting Independence, and Uncorrelated Choice. 
This, I seem to understand, was due to the erroneous belief that the three hypotheses are implied by 
and imply the impossibility of action-at-a-distance. 
To make the situation worse, this special usage of \lq\lq{}locality\rq\rq{} has variances: 
some authors refer to Setting Independence as the no-signaling hypothesis (erroneously, since if $B$ can change 
the local parameters $\lambda_B$, a message can be instantaneously sent to $A$), 
and to Reducibility of Correlations as a consequence of \lq\lq{}locality\rq\rq{}; 
other authors refer to both Setting Independence and Reducibility of Correlations as \lq\lq{}locality\rq\rq{} conditions, others still 
take locality as synonymous with no-signaling.
Throughout this paper we use locality as the impossibility of action-at-a-distance (see Appendix \ref{app:locdef}). 

\subsection{Leggett inequality}
Ref.~\citeonline{Leggett2003} considered a class of hidden-variable models which obey 
Uncorrelated Choice, as expressed by Eq.~\eqref{eq:uc}, and do not necessarily satisfy
Reducibility of Correlations, but which obey an analogue of Malus's law for the hidden variables, 
which consist in two unit vectors, $\lambda=\{\mathbf{u},\mathbf{v}\}$: 
\begin{subequations}\label{eq:malus}
\begin{align}
\label{eq:malusa}
M^A_\sigma(\lambda,\mathbf{a},\mathbf{b})=&(1+\sigma\mathbf{u}\cdot\mathbf{a})/2, \\
\label{eq:malusb}
M^B_\tau(\lambda,\mathbf{b},\mathbf{a})=&(1+\tau\mathbf{v}\cdot\mathbf{b})/2.
\end{align}
\end{subequations}
 It was shown\cite{Leggett2003,Groblacher2007,Branciard2008} that these models 
predicts a correlator satisfying an inequality known as Leggett inequality, which is violated by quantum mechanics and 
by experiments.\cite{Groblacher2007,Paterek2007,Branciard2007,Eisaman2008,Romero2010,Paternostro2010,Lee2011}  
It is unfortunately confusing that the new Leggett inequality was presented as an incompatibility theorem between 
quantum mechanics and a class of non-local models. 
Indeed, models satisfying Eqs.~\eqref{eq:malus} satisfy locality, 
if one thinks of $\mathbf{u}$ and $\mathbf{v}$ as parameters 
attached to each particle, while, as we discussed above, models satisfying Setting Independence in general do not. 
In this sense, Bell inequalities rule out also some non-local models, while Leggett inequality does not, contrary to what is 
claimed. 
The source of the misunderstanding is the following: 
in order to justify the correlation 
\begin{equation}
P_{\sigma,\tau}(\lambda,\mathbf{a},\mathbf{b})\neq 
M^A_\sigma(\lambda,\mathbf{a},\mathbf{b})M^B_\tau(\lambda,\mathbf{b},\mathbf{a}),
\end{equation}
Ref.~\citeonline{Leggett2003} assumed the existence of further variables $\lambda\rq{}$, in terms of which 
\begin{equation}
P_{\sigma,\tau}(\lambda,\mathbf{a},\mathbf{b})
=\int d\mu(\lambda\rq{}|\lambda) 
P_{\sigma,\tau}(\lambda,\lambda\rq{},\mathbf{a},\mathbf{b}),
\end{equation}
where the joint conditional probability is assumed to factorize as 
\begin{equation}
P_{\sigma,\tau}(\lambda,\lambda\rq{},\mathbf{a},\mathbf{b}) = 
\delta_{\sigma,S(\lambda,\lambda\rq{},\mathbf{a},\mathbf{b})}
\delta_{\tau,T(\lambda,\lambda\rq{},\mathbf{b},\mathbf{a})}.
\end{equation}
These models are non-local, since they violate Setting Independence, and deterministic. 
Leggett coined the term \lq\lq{}crypto-non-local\rq\rq{} theories to refer to them. 
Yet, it was realized that these models are only a special case of non-local models,\cite{Aspect2007} and that 
it is irrelevant that $\lambda\rq{}$ exist or not, given that Eqs.~\eqref{eq:uc} and \eqref{eq:malus} imply 
Leggett inequality, independently of how they are arrived to.\cite{Branciard2008,DiLorenzo2011b}
Furthermore, one could say as well that the violation of Bell inequality rules out all the crypto-non-local theories such that 
upon integration of $\lambda\rq{}$ Eqs.~\eqref{eq:factor}, \eqref{eq:si}, and \eqref{eq:uc} hold.
Thus, the importance of Leggett inequality consists not in the fact that it discriminates between quantum mechanics 
and non-local theories, but in that it does not require Reducibility of Correlations, just Uncorrelated Choice and 
a more specific form of Setting Independence. 

\subsection{Models reproducing the quantum predictions}
Examples that the violation of Uncorrelated Choice could lead to reproducing the quantum mechanical prediction were  provided in Ref.~\citeonline{Brans1988,Hall2010}. These models do not satisfy Malus\rq{}s law. 
The amount of violation of Uncorrelated Choice necessary to reproduce quantum mechanics 
was recently quantified.\cite{Hall2010,Barrett2011} 
In Section \ref{sec:model}, we discuss a model of ours\cite{DiLorenzo2012b} that satisfies at the same time 
the hypotheses of Malus\rq{}s law (and hence Setting Independence) and Reducibility of Correlations, but violates  
Uncorrelated Choice. 
In the literature there are also models violating both Uncorrelated Choice and 
Reducibility of Correlations,\cite{Cerf2005,Groblacher2007,Brunner2008} 
only Setting Independence,\cite{Toner2003,Pawlowski2010} or only Reducibility of Correlations.\cite{DiLorenzo2011g}   

Clearly, by violating one of the hypotheses at the basis of Bell and Leggett inequalities it may 
be possible to violate them. Reproducing quantum mechanics, however, is not guaranteed, 
since the violation of Bell and Leggett inequalities is a necessary 
but not sufficient condition. 

\section{Alternative derivation of Bell inequalities}\label{sec:bellboole}
More than a hundred years before Bell derived the famous inequality, Boole\cite{Boole1862} defined what he called \lq{}conditions of possible experience\rq{}:  
He imposed that the observed probabilities of $k$ events $E_{j_1},E_{j_2},\cdots E_{j_k}$ selected from a set $\{E_1,\cdots E_n\}$ were the marginals of a master probability $P(E_1,\cdots E_n)$. This led to inequalities of the form 
\begin{equation}
\sum_{r=1}^R A_r p_r \le A_0
\end{equation}
where $R$ is the number of different subsets of events for which there are experimental data, while $p_r$ is the corresponding probability, and 
$A_r$ are properly chosen coefficients. 
It is possible to derive Bell-type inequalities from Boole inequalities, as was done in various papers predating 
Bell\rq{}s\cite{Bass1955,Schell1955,Vorobev1959,Vorobev1962}, where Boole\rq{}s inequalities were rediscovered 
independently. These papers, however, were concerned with mathematical and computational problems, and their authors 
were not aware of the implications of the results in physics.  
As an example, the CHSH inequality can be derived in a few lines:
\begin{align}
\nonumber
&|C(\mathbf{a},\mathbf{b})+C(\mathbf{a}\rq{},\mathbf{b})+C(\mathbf{a},\mathbf{b}\rq{})-C(\mathbf{a}\rq{},\mathbf{b}\rq{})|
\\
\nonumber
&=\bigg|\sum_{\substack{\sigma,\sigma\rq{}\\{\tau,\tau\rq{}}}} \left(\sigma \tau+\sigma\rq{} \tau+\sigma \tau\rq{}-\sigma\rq{} \tau\rq{}\right)
P_{\sigma,\tau,\sigma\rq{},\tau\rq{}}(\mathbf{a},\mathbf{b},\mathbf{a}\rq{},\mathbf{b}\rq{})\bigg|\\
\nonumber
&\le \sum_{\substack{\sigma,\sigma\rq{}\\{\tau,\tau\rq{}}}} |\sigma \tau+\sigma\rq{} \tau+\sigma \tau\rq{}-\sigma\rq{} \tau\rq{}|
\, P_{\sigma,\tau,\sigma\rq{},\tau\rq{}}(\mathbf{a},\mathbf{b},\mathbf{a}\rq{},\mathbf{b}\rq{})\\
&\le \sum_{\substack{\sigma,\sigma\rq{}\\{\tau,\tau\rq{}}}}\left( |\sigma +\sigma\rq{}|+|\sigma -\sigma\rq{} |\right)
P_{\sigma,\tau,\sigma\rq{},\tau\rq{}}(\mathbf{a},\mathbf{b},\mathbf{a}\rq{},\mathbf{b}\rq{})= 2 .
\end{align}
The beauty of this alternative route to Bell-type inequalities consists in that it does not rely on the 
existence of hidden variables and requires only one hypothesis, 
the \lq{}condition of possible experience\rq{}, or, in current terminology, counter-factual definiteness: 
spin components along any direction have a well defined probability distribution, independently of which 
component is being measured. This is but the definition of \lq{}elements of reality\rq{} given in the paper 
by Einstein, Podolsky, and Rosen.\cite{Einstein1935} 
While it is dubious that the three hypotheses that Bell formulated in his derivation of the inequality 
and the very concept of hidden variables correspond to the hypotheses made in the EPR paper, counter-factual definiteness, instead, is certainly among the latter ones. 
Thus, the Boole-style\rq{}s derivation of the Bell-type inequalities, and the experimental violation of these, proves 
the EPR assumption of \lq{}elements of reality\rq{} fallacious. 
This was indeed understood immediately, albeit in a qualitative way, in many 
responses\cite{Kemble1935,Ruark1935,Bohr1935,Wolfe1936,Furry1936a,Furry1936b} to the EPR paper, of 
which Bohr\rq{}s\cite{Bohr1935} is the most famous, but perhaps not the most clear. 

I conclude this section acknowledging 
Refs.~\citeonline{Pitowsky1994} and \citeonline{Hess2005} that brought to the attention of the physics community the 
pioneering results of Boole and those of Bass, Schell, and Vorob\rq{}ev, respectively. 
Furthermore, the Boole-style derivation of the inequalities was rediscovered independently in 
Refs.~\citeonline{Fine1982a} and \citeonline{Accardi2001}. Finally, Ref.~\citeonline{Durt1997} proved that the observed frequencies of experiments (the frequency of observing, e.g., an outcome $\sigma$ times the frequency with which the detector is chosen along a direction $\mathbf{a}$) always admit a 
master probability distribution, i.e., they are Kolmogovorian in the terminology of Ref.~\citeonline{Durt1997}). 
This shows quantitatively that superdeterminism can never be ruled out.

\section{The Toner and Bacon model}\label{sec:tb}
In order to show an example of the interchangeability between non-locality and limited free will, let us consider the Toner and Bacon model,\cite{Toner2003} 
which relies on a station sending a bit of information $c$ to the other one, presumably by instantaneous communication, otherwise one could not violate Bell inequality in real time 
without keeping the locality loophole open. The model works in the following way: 
both $A$ and $B$ can access two unit-vector hidden-variables $\mathbf{u},\mathbf{v}$, which have a uniform distribution $\mu_{\mathbf{u},\mathbf{v}}=1/(4\pi)^2$ independent 
of the settings. The outcome of $A$ is deterministic, its marginal probability being 
\begin{equation}\label{eq:tbmarg}
M^A_\sigma(\mathbf{a},\mathbf{u},\mathbf{v})=
\delta_{\sigma,S(\lambda,\mathbf{a})}
\end{equation}
with $S(\lambda,\mathbf{a})=\mathrm{sgn}(\mathbf{u}\cdot\mathbf{a})$, and 
the sign function $\mathrm{sgn}(x)=-1, x<0$ and $\mathrm{sgn}(x)=+1, x>0$. 
The conditional probability at $B$ is 
\begin{equation}\label{eq:tbcond}
Q^B_\tau(\mathbf{b},\mathbf{u},\mathbf{v},\mathbf{a},\sigma)=
\begin{cases}
\frac{0}{0} \ , \ &\text{ if } \sigma= -S(\lambda,\mathbf{a})\\
\delta_{\tau,T(\lambda,\mathbf{a},\mathbf{b})}
\ , \ &\text{ if } \sigma= S(\lambda,\mathbf{a})
\end{cases}
\end{equation}
where the indeterminate form $\tfrac{0}{0}$ reflects the fact that the priors $\mathbf{u},\mathbf{a},\sigma$ are incompatible, 
$T(\lambda,\mathbf{a},\mathbf{b})=-\mathrm{sgn}[(\mathbf{u}+c(\lambda,\mathbf{a})\mathbf{v})\cdot\mathbf{b}]$, 
$c(\lambda,\mathbf{a})=\mathrm{sgn}(\mathbf{u}\cdot\mathbf{a})\,\mathrm{sgn}(\mathbf{v}\cdot\mathbf{a})$. 
\subsection{Two possible extensions of the Toner and Bacon model.}
Let us lift the hypothesis of determinism so that, instead of Eq.~\eqref{eq:tbmarg}, we have 
\begin{equation}\label{eq:tbmarg2}
M^A_\sigma(\mathbf{a},\mathbf{u},\mathbf{v})=
p\, \delta_{\sigma,S(\lambda,\mathbf{a})}+(1-p)\delta_{\sigma,-S(\lambda,\mathbf{a})} ,
\end{equation}
where $0<p<1$ is a fixed probability. 
There is an ambiguity about how to interpret the bit $c$. 
The latter, indeed, can be seen either as a function of $\mathbf{u},\mathbf{v},\mathbf{a}$, or, equivalently, 
as $c=\tilde{c}(\mathbf{v},\mathbf{a},\sigma)=\sigma\mathrm{sgn}(\mathbf{v}\cdot\mathbf{a})$.
Then Eq.~\eqref{eq:tbcond} becomes either
\begin{subequations}
\begin{equation}\label{eq:tbcond2}
Q^B_\tau(\mathbf{b},\mathbf{u},\mathbf{v},\mathbf{a},\sigma)=
\delta_{\tau,T(\lambda,\mathbf{a},\mathbf{b})}
 \ , 
\end{equation}
if we choose $c=c(\lambda,\mathbf{a})$, 
or
\begin{equation}\label{eq:tbcond3}
Q^B_\tau(\mathbf{b},\mathbf{u},\mathbf{v},\mathbf{a},\sigma)=
\delta_{\tau,\tilde{T}(\lambda,\mathbf{a},\mathbf{b},\sigma)},
\end{equation}
\end{subequations}
with $\tilde{T}(\lambda,\mathbf{a},\mathbf{b},\sigma)=
-\mathrm{sgn}[(\mathbf{u}+\tilde{c}(\mathbf{v},\mathbf{a},\sigma)\mathbf{v})\cdot\mathbf{b}]$, 
if we choose $c=\tilde{c}(\mathbf{v},\mathbf{a},\sigma)$. 
We have thus two families of hidden variable models, depending on a parameter $p$. 
In the first family, the conditional probability does not depend on $\sigma$, and it coincides with the marginal probability, 
which, hence, depends on $\mathbf{a}$, violating Setting Independence. 
After integrating out $\mathbf{u},\mathbf{v}$, the probability is 
\begin{subequations}
\begin{equation}\label{eq:tbout1}
P_{\sigma,\tau}(\mathbf{a},\mathbf{b}) = \frac{1}{4}\left[1-(2p-1)\sigma\tau \mathbf{a}\cdot\mathbf{b}\right]. 
\end{equation}
In the second family, we have that Reducibility of Correlations is violated, instead, while the marginal probability at station 
$A$ satisfies Setting Independence. 
Since the Toner and Bacon model has the (unaesthetic, in my opinion) feature of being asymmetric, 
if we instead consider the marginal probability at $B$ and the conditional probability at $A$, we arrive to 
\begin{equation*}
M^B_\tau(\mathbf{b},\mathbf{u},\mathbf{v},\mathbf{a})=
p\delta_{\tau,\tilde{T}(\lambda,\mathbf{a},\mathbf{b},S(\mathbf{u},\mathbf{a}))}
+(1-p)\delta_{\tau,\tilde{T}(\lambda,\mathbf{a},\mathbf{b},-S(\mathbf{u},\mathbf{a}))},
\end{equation*}
and
\begin{equation*}
Q^A_\sigma(\mathbf{a},\mathbf{u},\mathbf{v},\mathbf{b},\tau)=
\frac{p\, \delta_{\sigma,S(\lambda,\mathbf{a})}
\delta_{\tau,\tilde{T}(\lambda,\mathbf{a},\mathbf{b},S(\mathbf{u},\mathbf{a}))}
+(1-p)\delta_{\sigma,-S(\lambda,\mathbf{a})}\delta_{\tau,\tilde{T}(\lambda,\mathbf{a},\mathbf{b},-S(\mathbf{u},\mathbf{a}))}}{M^B_\tau(\mathbf{b},\mathbf{u},\mathbf{v},\mathbf{a})},
\end{equation*}
so that both Setting Independence and Reducibility of Correlations are violated by the second family in this alternative 
factorization. 
However, for asymmetric models, we propose to say that Setting Independence is violated when it is 
not satisfied in either of the two possible factorizations of the joint probability, 
$P_{\sigma,\tau}(\lambda,\mathbf{a},\mathbf{b})=
M^{A}_{\sigma}(\lambda,\mathbf{a},\mathbf{b}) 
Q^B_{\tau}(\lambda,\mathbf{b},\mathbf{a},\sigma)=
Q^{A}_{\sigma}(\lambda,\mathbf{a},\mathbf{b},\tau) 
M^B_{\tau}(\lambda,\mathbf{b},\mathbf{a})$. 
(If Reducibility of Correlations is satisfied in one factorization, it is also in the alternative one.)
Therefore, in this second case, Setting Independence is satisfied. 
The resulting average probability is 
\begin{equation}\label{eq:tbout2}
P_{\sigma,\tau}(\mathbf{a},\mathbf{b}) = \frac{1}{4}\left[1-p\sigma\tau \mathbf{a}\cdot\mathbf{b}\right]. 
\end{equation}
\end{subequations}
Both Eq. \eqref{eq:tbout1} and \eqref{eq:tbout2} reproduce the quantum mechanical probability for $p\to 1$, and both 
families of models tend to the Toner and Bacon model in this limit: 
the first family violates Setting Independence and 
preserves Reducibility of Correlations, the second family does the opposite. 
In a sense, models not complying with the Reducibility of Correlations are more robust, 
since they deviate from the quantum mechanical probability less than the models not complying with 
Setting Independence (correlations are reduced by $p$ instead than by $2p-1$).
\subsection{Local realization through limited free will}
Finally, let us show that the Toner and Bacon model admits a simple reinterpretation in terms of limited free will. 
We assume that $c$ is not a bit generated by $A$, but a binary hidden-variable, in addition to 
two unit vectors $\mathbf{u},\mathbf{v}$, which are distributed uniformly. 

When $A$ receives $\mathbf{u},\mathbf{v},c$, the setting will be chosen so that $c=c(\mathbf{u},\mathbf{v},\mathbf{a})$ 
holds. We may imagine that $\sigma=S(\mathbf{u},a)$ is enforced by some appropriate local physical law. 
Analogously, $B$ chooses an orientation $\mathbf{b}$, but freely, while the physical laws on his side determine that 
$\tau=T(\mathbf{u},\mathbf{v},c,\mathbf{b})=S(-\mathbf{u}-c\mathbf{v},\mathbf{b})$. 
Hence, the hidden variables are, in this reformulation, $\lambda=\{\mathbf{u},\mathbf{v},c\}$, and they are 
distributed according to
\begin{equation}\label{eq:tbhv1}
\mu\rq{}_{\mathbf{u},\mathbf{v},c}(\mathbf{a},\mathbf{b})=
\frac{1}{(4\pi)^2}\,\delta_{c,c(\mathbf{u},\mathbf{v},\mathbf{a})} .
\end{equation}
The marginal probability at $A$ is still 
\begin{equation}\label{eq:tbmarg1}
M^A_\sigma(\mathbf{u},\mathbf{v},c,\mathbf{a})=
\delta_{\sigma,S(\mathbf{u},\mathbf{a})}
\end{equation}
while the conditional probability at $B$ is 
\begin{equation}\label{eq:tbcond1}
Q^B_\tau(\mathbf{b},\mathbf{u},\mathbf{v},\mathbf{a},\sigma)=
\begin{cases}
\frac{0}{0} \ , \ &\text{ if } \sigma= -S(\mathbf{u},\mathbf{a})\\
\delta_{\tau,T\rq{}(\lambda,\mathbf{b})}
\ , \ &\text{ if } \sigma= S(\mathbf{u},\mathbf{a})
\end{cases}
\end{equation}
with $T\rq{}(\lambda,\mathbf{b})=-\mathrm{sgn}[(\mathbf{u}+c\mathbf{v})\cdot\mathbf{b}]$, 
so that both Setting Independence and Reducibility of Correlations are satisfied.

\section{Hall model}\label{sec:hall}
Ref.~\citeonline{Hall2010} proposes a different decomposition of the quantum mechanical probability, through a model 
which has the maximum ``free will'' compatible with quantum mechanics, and deterministic outcomes: 
\begin{equation}\label{eq:halldens}
\mu(\mathbf{u},\mathbf{v}|\mathbf{a},\mathbf{b})=\delta(\mathbf{u}+\mathbf{v}) 
\frac{1-f(\mathbf{u},\mathbf{v}, \mathbf{a},\mathbf{b})}{8\arccos{f(\mathbf{u},\mathbf{v}, \mathbf{a},\mathbf{b})}},
\end{equation}
where 
\begin{equation}\label{eq:hallf}
f(\mathbf{u},\mathbf{v}, \mathbf{a},\mathbf{b})=
\mathrm{sgn}(\mathbf{u}\cdot\mathbf{a})
\mathrm{sgn}(\mathbf{v}\cdot\mathbf{b})\ \mathbf{a}\cdot\mathbf{b} ,
\end{equation}
and $\mathrm{sgn}(x)=1$ for $x\ge 0$, $\mathrm{sgn}(x)=-1$ for $x< 0$.  
The probability of the outcomes, given $\mathbf{u},\mathbf{v}$, is 
\begin{equation}\label{eq:hallprob}
P_{\sigma,\tau}(\mathbf{u},\mathbf{v}, \mathbf{a},\mathbf{b})=
\delta_{\sigma,\mathrm{sgn}(\mathbf{u}\cdot\mathbf{a})} \delta_{\tau,\mathrm{sgn}(\mathbf{v}\cdot\mathbf{b})},
\end{equation}
so that both Setting Independence and Reducibility of Correlations are satisfied. 
We notice that if $\mathbf{u},\mathbf{v}$ are considered local parameters, localized, respectively, at 
the particle reaching $A$ and the one reaching $B$, the model is non-local, because 
of the tangled nature of the function $f$. 
Indeed, the marginal distribution for $\mathbf{u}$ is 
\begin{equation}
\mu^A(\mathbf{u}|\mathbf{a},\mathbf{b})= 
\frac{1-f(\mathbf{u},-\mathbf{u}, \mathbf{a},\mathbf{b})}{8\arccos{f(\mathbf{u},-\mathbf{u}, \mathbf{a},\mathbf{b})}},
\end{equation}
and clearly depends on $\mathbf{b}$. Thus, the two vectors are to be considered global parameters, if one 
wants to interpret Eq.~\eqref{eq:halldens} as limiting the \lq\lq{}free will\rq\rq{} of the observers who choose
$\mathbf{a},\mathbf{b}$ but preserving locality. 
Assuming that the unconditional probabilities are uniformly distributed, 
by using Bayes\rq{}s theorem one finds the conditional probability of having setting $\mathbf{a},\mathbf{b}$ for 
given $\mathbf{u}$, 
\begin{equation}
\Pi(\mathbf{a},\mathbf{b}|\mathbf{u}) = \frac{1-f(\mathbf{u},-\mathbf{u}, \mathbf{a},\mathbf{b})}{32\pi\arccos{f(\mathbf{u},-\mathbf{u}, \mathbf{a},\mathbf{b})}}.
\end{equation}
Thus not only the settings are not free, but they are also correlated. 
In Appendix \ref{app:real}, we show how to realize this probability distribution by using two additional global unit vectors 
$\mathbf{z}_A,\mathbf{z}_B$, in terms of which the distribution is 
\begin{equation}\label{eq:halldens2}
\mu\rq{}(\mathbf{u},\mathbf{v},\mathbf{z}_A,\mathbf{z}_B|\mathbf{a},\mathbf{b})=
\delta(\mathbf{u}+\mathbf{v}) \delta(\mathbf{a}-\mathbf{z}_A) 
\delta(\mathbf{b}-\mathbf{z}_B) 
\frac{1-f(\mathbf{u},\mathbf{v}, \mathbf{z}_A,\mathbf{z}_B)}{8\arccos{f(\mathbf{u},\mathbf{v}, \mathbf{z}_A,\mathbf{z}_B)}}.
\end{equation}
Then if $\mathbf{u},\mathbf{v}$ are interpreted as local parameters 
the marginal distribution becomes 
\begin{equation}
\mu^A(\mathbf{u}|\mathbf{z}_A,\mathbf{z}_B,\mathbf{a},\mathbf{b})= 
\frac{1-f(\mathbf{u},-\mathbf{u}, \mathbf{z}_A,\mathbf{z}_B)}{8
\arccos{f(\mathbf{u},-\mathbf{u}, \mathbf{z}_A,\mathbf{z}_B)}},
\end{equation}
for $\mathbf{a}=\mathbf{z}_A$, $\mathbf{b}=\mathbf{z}_B$, and is undeterminate in the other cases. 
The hypothesis of locality is satisfied by this trivial introduction of additional global parameters. 
We notice that it is important to distinguish the variables from the values they take. 
The $\delta$ function in Eq.~\eqref{eq:halldens2} 
force the value of $\mathbf{a}$ to equal the value of $\mathbf{z}_A$, etc., but $\mathbf{a}$ and $\mathbf{z}_A$ 
are distinct variables.

\section{A simple model reproducing the quantum mechanical probabilities for the spin singlet}\label{sec:model}
Here we present a simple model reproducing the quantum mechanical probability for a spin-singlet. 
The model satisfies at the same time the hypotheses of Setting Independence, Reducibility of Correlations, and Compliance with Malus\rq{}s law, 
while it violates the hypothesis of Uncorrelated Choice. 
This model was derived already in 2007, but it remained unpublished until recently,\cite{DiLorenzo2012b} 
when I came to realize that despite its simplicity it has 
several appealing features. In the basic formulation, the hidden-variables consist in two unit vectors $\mathbf{u},\mathbf{v}$, 
localized at the left- and right-going particle, respectively (see Fig. \ref{fig:eprbsetup}). As one would expect for a state with zero total spin, 
the vectors are perfectly anti-correlated $\mathbf{u}=\mathbf{u}_0, \mathbf{v}=-\mathbf{u}_0$ 
(an index is used for the values, while the random variables are indicated by letters with no index). 
Furthermore, the distribution of $\mathbf{u}$ and $\mathbf{v}$ is tied to the polarization settings 
$\mathbf{a},\mathbf{b}$ according to 
\begin{align}
\nonumber
\mu_{\mathbf{u}=\mathbf{u_0},\mathbf{v}=\mathbf{v}_0}(\mathbf{a},\mathbf{b})=&
\frac{1}{4}\delta(\mathbf{u}_0+\mathbf{v}_0)\\ 
\label{eq:hvdistr}
&\times \left[\delta(\mathbf{a}-\mathbf{u}_0)+\delta(\mathbf{a}+\mathbf{u}_0)+
\delta(\mathbf{b}-\mathbf{v}_0)+\delta(\mathbf{b}+\mathbf{v}_0)\right].
\end{align}
The marginal probability of a particle with $\mathbf{u}=\mathbf{u}_0$ giving an output $\sigma$ when impinging on a detector 
with polarization $\mathbf{a}$ is given by Malus\rq{}s law (which is local: no action-at-a-distance is needed)
\begin{equation}\label{eq:margprob}
M^A_\sigma(\mathbf{u},\mathbf{a}) = \frac{1}{2}\left[1+\sigma \mathbf{u}\cdot\mathbf{a}\right].
\end{equation}
At the hidden variable level, the particles are uncorrelated, i.e. the two vectors $\mathbf{u},\mathbf{v}$ specify all needed information, 
so that the conditional probability coincides with the marginal one, 
\begin{equation}\label{eq:condprob}
Q^B_\tau(\mathbf{v},\mathbf{b},\mathbf{a},\sigma) = M^B_\tau(\mathbf{v},\mathbf{b})=\frac{1}{2}\left[1+\tau \mathbf{v}\cdot\mathbf{b}\right].
\end{equation}
After integration over $\mathbf{u}_0,\mathbf{v}_0$ the quantum mechanical probability for the singlet is recovered 
\begin{equation}\label{eq:singletprob}
P_{\sigma,\tau}(\mathbf{a},\mathbf{b})=\frac{1}{4}\left[1-\sigma \tau \mathbf{a}\cdot\mathbf{b}\right] .
\end{equation}
Equations \eqref{eq:margprob} an \eqref{eq:condprob} are classical, i.e., they can be realized through classical resources. 
By demonstrating that Eq.~\eqref{eq:hvdistr} can be realized through classical resources as well, the main result of this paper 
will be achieved: so-called quantum correlations are not exclusively quantum, since they can be mimicked by classical means. 
We remark that the interpretation of Eq.~\eqref{eq:hvdistr}, or of any mathematical expression for probabilities, relies on 
a metaphysical assumption, the one about cause and effect.  As will be discussed in Section \ref{sec:inter}, by interchanging cause and effect 
what was attributed to lack of \lq\lq{}free will\rq\rq{} is attributed to non-local action-at-a-distance, and vice versa. 

\subsection{Local realization through one bit of shared randomness between the measuring stations}\label{subsec:model} 
Let us discuss how to realize Eq.~\eqref{eq:hvdistr} assuming $\mathbf{u},\mathbf{v}$ to be the cause, 
and $\mathbf{a},\mathbf{b}$ the effect. 
The two parties $A$ and $B$ possess identical pseudo-random\footnote{By making use of classical resources, the only known 
way to produce two identical numbers at spacelike separated locations is to rely on a pseudo-random algorithm using an identical seed.}  
number generators, each producing the same binary number $c\in\{0,1\}$. 
Furthermore, by making a local measurement, they can determine, respectively, $\mathbf{u}$ and $\mathbf{v}$ (see Fig. \ref{fig:eprbsetup}) without perturbing the system. 
When $c=0$, $A$ will measure $\mathbf{u}$, then flip a coin, and according to the outcome $d$  of the flip (or at whim), 
will orient the polarizer either in direction $\mathbf{u}_0$ or $-\mathbf{u}_0$, obtaining the outcome $\sigma=+1$ in the first case, $\sigma=-1$ in the second case. 
$B$, on the other hand, shall choose $\mathbf{b}$ freely (for instance at random) when $c=0$. 
In the other case $c=1$, $A$ and $B$ will reverse their actions, i.e. $A$ shall choose $\mathbf{a}$ freely, while $B$ chooses $\mathbf{b}=\mathbf{v}_0$ or  
$\mathbf{b}=-\mathbf{v}_0$ according to the flipping of a coin (or at whim). 
Given that the conservation law $\mathbf{u}+\mathbf{v}=0$ holds, Eq.~\eqref{eq:hvdistr} is obtained after summing 
over $c$ and $d$, which do not appear in the conditional probability $P_{\sigma,\tau}(\mathbf{u},\mathbf{v},\mathbf{a},\mathbf{b})$ the distribution that we have just described 
\begin{align}
\mu_{\lambda_0}(\mathbf{a},\mathbf{b})=&
\frac{1}{8}\delta(\mathbf{u}_0+\mathbf{v}_0) \delta\biggl(\overline{c_0}(\mathbf{a}-d_0^{(A)}\mathbf{u}_0)+c_0(\mathbf{b}-d_0^{(B)}\mathbf{v}_0)\biggr) ,
\label{eq:hvdistr2}
\end{align}
with $c\in\{0,1\}$, $\overline{c}=1-c$, and $d\in\{Heads, Tails\}=\{-1,1\}$.\footnote{Eq.~\eqref{eq:hvdistr} may also be realized in 
an asymmetric way, e.g., having $A$ to always orient the polarizer along $\pm\mathbf{u}$.}
Some authors may refer to the bit $c$ as \lq\lq{}non-local information\rq\rq{}, but this is evidently just a verbal trick, since no 
instantaneous communication is going on, nor there is any action-at-a-distance.  

It is interesting to compare two different proposed measures of \lq\lq{}free will\rq\rq{} in relation to this model. 
According to Ref.~\citeonline{Hall2010}, putting $\lambda=\{\mathbf{u},\mathbf{v},c,d\}$
\begin{equation}
M=\mathrm{sup}\sum_{c,d}\int d\mathbf{u}d\mathbf{v} 
\left|\,\mu_\lambda(\mathbf{a},\mathbf{b})-\mu_\lambda(\mathbf{a}',\mathbf{b}')\right|=2 .
\end{equation}
This is the maximum value of $M$ and corresponds to the absence of ``free will''. 
However, since in our model half of the times $A$ or $B$ is free to choose the setting, while $B$ or $A$, respectively, has still a binary choice, 
$M$ does not quantify sufficiently well the concept of \lq\lq{}free will\rq\rq{}. 
The mutual information proposed in Ref.~\citeonline{Barrett2011}, gives instead  
$I(\mathbf{a},\mathbf{b}:\mathbf{u},\mathbf{v},c,d)=\infty$. 
However, if we discretize the allowed polarizations, having them take $N$ possible values, 
we get $I(\mathbf{a},\mathbf{b}:\mathbf{u},\mathbf{v},c,d)=\log_2(N)$, half of the maximum value $I_{max}(\mathbf{a},\mathbf{b}:\mathbf{u},\mathbf{v},c,d)=2\log_2(N)$, 
which represents total absence of free will. 
This measure is hence a better quantifier of \lq\lq{}free will\rq\rq{}.
\subsection{Local realization exploiting the detection loophole.}
As pointed in Ref.~\citeonline{Barrett2011}, there is a close relation between the violation of 
Uncorrelated Choice and the detection loophole (which consists in the fact that not all pairs of particles impinging 
on the detectors successfully give rise to coincidence counts). 
This was explored in Ref.~\citeonline{Gerhardt2011} to violate a Bell inequality (with only two possible settings for each detector).  
Indeed, an alternative explanation of the probability of the former subsection is that 
each particle carries with it a bit $c_A$ (resp., $c_B$) and a unit vector $\mathbf{u}$ (resp., $\mathbf{v}$), where the variables are anticorrelated $c_A+c_B=1\, (\text{modulo 2})$, $\mathbf{u}+\mathbf{v}=0$. 
The particles are programmed to act according to the following set of instructions: 
If $c_A=0$, particle $A$ shall behave according to Malus\rq{}s law, giving an output $\sigma$ 
with the probability given in Eq.~\eqref{eq:malusa}. 
If $c_A=1$, particle $A$ will give no output unless the direction $\mathbf{a}$ coincides (let us say within some tolerance, 
so that we can avoid continuous distributions) with $\pm\mathbf{u}$. 
Particle $B$ acts in a corresponding way. 
If one is interested only in violating a Bell inequality, one needs two settings for $\mathbf{a}$ and two for $\mathbf{b}$, 
so that one may limit $\mathbf{u}$ to take eight possible values, and obtain this way an efficiency 
of $25\%$ (i.e. $75\%$ of times only one detector will give an output).  Also, if one makes the model asymmetric, having, e.g., always $c_A=0$, then 
the detection efficiency drops only to $50\%$, as in Ref.~\citeonline{Gerhardt2011}. 
However, if $\mathbf{a},\mathbf{b}$ have an uncertainty $\Delta \Omega$ and they are spanned throughout 
the unit sphere, the protocol illustrated here reduces the efficiency to $\Delta \Omega/2\pi$. 

\subsection{Local realization of the model through shared randomness between each station and the entangler}\label{subsec:real}
There is no need to have 
shared randomness between $A$ and $B$, but only between each station and the entangler. 
This is illustrated at length in Appendix \ref{app:real}.
\subsection{Non-local realization through action-at-a-distance}
 Let us show, now, that inverting the cause-effect relationship between $\mathbf{a}, \mathbf{b}$ and $\lambda$, 
 the model proposed can be explained alternatively by non-local action at a distance. 
 Indeed, Eq.~\eqref{eq:hvdistr}, 
This could be interpreted as having, e.g., $\mathbf{a}$ to force the value 
$\mathbf{u}_0=\mathbf{a}$ and $\mathbf{v}_0=-\mathbf{a}$ when $c_0=0$ and $d_0^{(A)}=1$. 
Since $\mathbf{v}$ is assumed to be a local parameter describing the particle at station $B$, we have an instance of 
action-at-a-distance: the setting at one station influences the parameter at the remote station.

\section{A model violating Cirel\rq{}son bound.}\label{sec:cirel}
We show that it is possible to violate the Cirel'son bound, which establishes 
the maximum violation of Bell inequality that quantum mechanics can achieve ($2\sqrt{2}$, when the Bell limit is $2$). 
All we need to do is to mix the two models of Refs.~\citeonline{Hall2010} and \citeonline{DiLorenzo2012b}: 
The distribution of the hidden variables is given by Eq.~\eqref{eq:hvdistr} and 
the $\lambda$-conditioned probability is not given by Malus\rq{}s law, but by Eq.~\eqref{eq:hallprob}. 
This gives the joint probability 
\begin{align}\label{eq:mixprob}
P_{\sigma,\tau}(\mathbf{a},\mathbf{b})&
=\frac{1}{4}\left[1-\sigma\tau\mathrm{sgn}(\mathbf{a}\cdot\mathbf{b})\right] .
\end{align}
\begin{figure}
\centering
\includegraphics[width=4in]{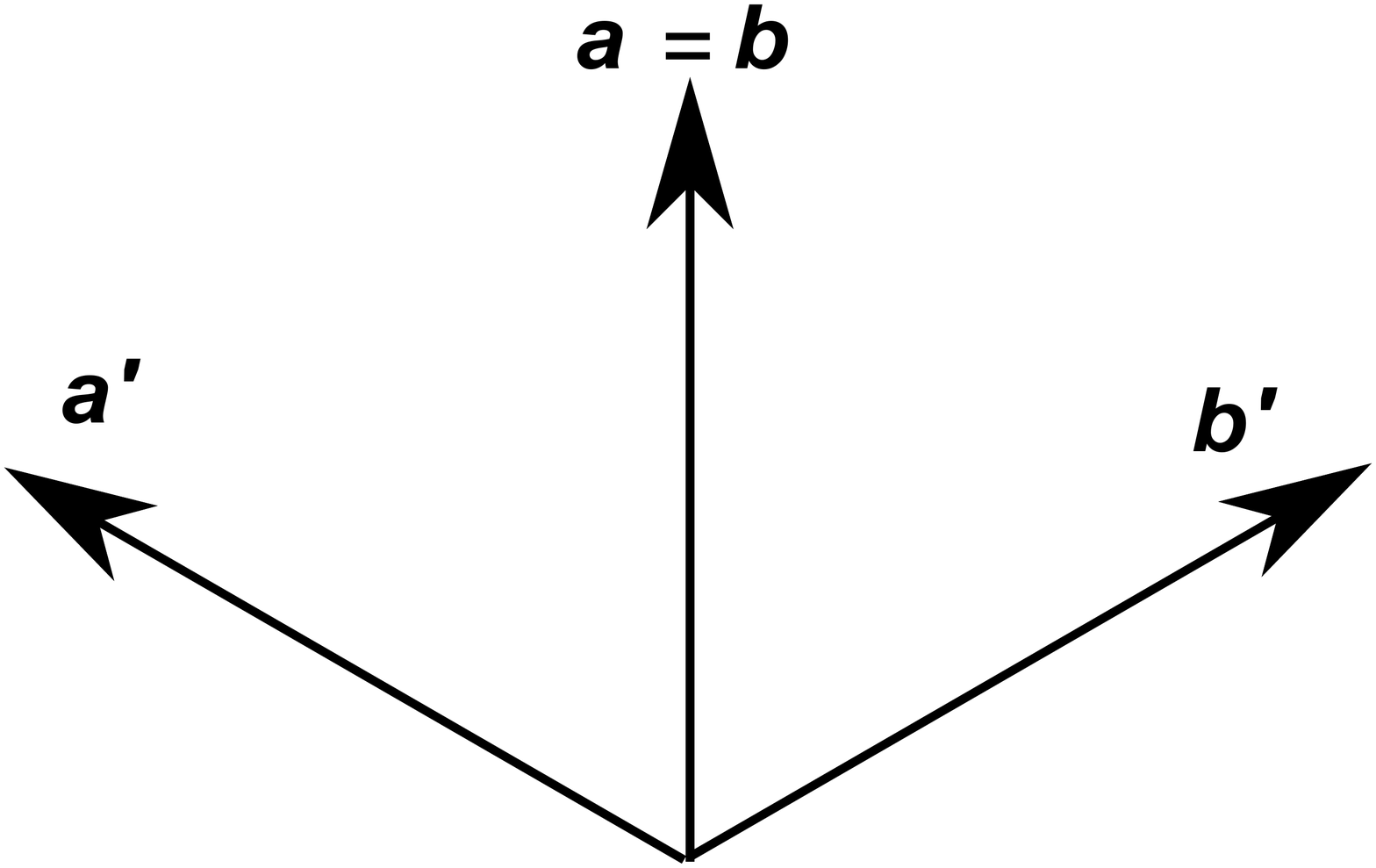}
\caption{\label{fig:maxbell} A possible configuration allowing $\mathcal{E}=4$ for the probability 
of Eq.~\eqref{eq:mixprob}.}
\end{figure}
The correlator is thus 
\begin{equation}
C(\mathbf{a},\mathbf{b})=-\mathrm{sgn}(\mathbf{a}\cdot\mathbf{b}) , 
\end{equation}
and the Clauser-Horne parameter 
$\mathcal{E}=|C(\mathbf{a},\mathbf{b})+C(\mathbf{a}',\mathbf{b})+C(\mathbf{a},\mathbf{b}')
-C(\mathbf{a}',\mathbf{b}')|$ 
reaches the value 4 for infinitely many choices of the orientations, one of which is given in Fig.~\ref{fig:maxbell}. 
This model can be realized with local classical resources along the lines of the previous section.

\section{How to discriminate among ``slave will", ``non-locality", and ``conspiracy"}\label{sec:inter}
The alternative realizations of the same model presented in Section \ref{sec:model} show that a probabilistic theory has several possible interpretations depending on the metatheoretical assumptions about the nature of its objects ad events. 
Let us see how one could decide in favor of one interpretation over the other. 
On one hand, the realization through the detection loophole can be falsified by making sure the detector\rq{}s efficiency exceeds $50\%$. 
The realization through limited free will may be discerned from the one through action-at-a-distance in the following way: 
Let us assume that $A$ and $B$ can measure the parameters $\mathbf{u},\mathbf{v}$ (if the parameters are not experimentally accessible or controllable, there is no point in assuming 
they exist) during the flight of the particles, before detecting $\sigma,\tau$. 
After several runs, $A$ will realize that half of the times $\mathbf{u}$ is directed along $\mathbf{a}$ or $-\mathbf{a}$, and that if she changes $\mathbf{a}$, $\mathbf{u}$ will change 
accordingly. 
Since the conservation law guarantees that $\mathbf{v}=-\mathbf{u}$, $A$, assuming she has free will, will conclude that she is acting instantaneously on the remote parameter. 
Then she will agree with $B$ the following instantaneous signaling protocol: they both limit their choices to four possible directions, $\pm \mathbf{a}$, $\pm \mathbf{b}$, e.g., orthogonal 
to each other. $A$ fixes her setting to $\mathbf{a}$ and $B$ to $\mathbf{b}$, then $A$ measures $\mathbf{u}$ and $B$ measures $\mathbf{v}$. 
If $A$ measures $\mathbf{u}=\mathbf{a}$, she will quickly switch her apparatus either to $\mathbf{b}$, making sure that $\mathbf{v}=-\mathbf{b}$ so that 
$B$ will measure with certainty $\tau=-1$ (which they agree to correspond to the bit 0), or to $-\mathbf{b}$, making sure that $\mathbf{v}=\mathbf{b}$ so that 
$B$ will measure with certainty $\tau=+1$ (which they agree to correspond to the bit 1). 
The half of the times when $\mathbf{u}=\pm\mathbf{b}$, $A$ will do nothing ($B$ knows that this is a fail, because he will see $\mathbf{v}=\mp\mathbf{b}$ before measuring $\tau$). 
This way, $A$ can send to $B$ an arbitrary sequence of bits. 
Now, if $A$ is actually acting with free will and there is an action-at-a-distance, then this sequence of bits will correspond to a significant message, and will have a low entropy. 
If instead $A$ has no free will, but it is the value of $\mathbf{u}$ that determined the setting $\mathbf{a}$, this sequence of bits will be totally random. 
Of course there is always the possibility of superdeterminism, where the same mechanism that determines $\mathbf{a},\mathbf{b}$ is also responsible 
for the thousand of years of evolution of the languages, and of the recent conventional encoding of their symbols in sequences of bits. This hypothesis is clearly unfalsifiable, 
i.e. we cannot set its probability of being true to exactly zero, but I believe that given our experience of the world and our scientific knowledge its probability must be extremely low.  

We notice that the slave will model cannot be tested unless one is able to measure or fix the additional parameters. 
For instance, Ref.~\citeonline{Scheidl2010} tests the free will hypothesis, but only for stochastic theories, and concedes that what is called therein \lq\lq{}deterministic local realism\rq\rq{} 
cannot be falsified. Indeed, it may happen that the random number generators and the hidden variables are correlated, but that any correlations and regularities disappear 
upon integrating out the hidden parameters. This would give the illusion of having two independent random number generators at the quantum level, while 
the randomness is lifted, or limited, at the subquantum level. Thus, for a given model that violates Uncorrelated Choice,  discriminating between slave will and 
non-locality requires the ability to either control or detect the hidden parameters. 

Finally, the realization of Section \ref{subsec:model} can be interpreted indifferently as stemming from a conspiracy or from a limitation of free will. 
Which is the case can be decided by making sure that neither $A$ nor $B$ can decide $\mathbf{a},\mathbf{b}$ based on a measurement of $\mathbf{u}, \mathbf{v}$. 
For instance, one could ask $A$ and $B$ to provide beforehand a list of the settings they will be using in each repetition. 
The procedure is opposite to the one followed to close the locality loophole: the settings are decided well in advance of the production of the entangled pairs, and they are 
kept fixed. If one observes $A$ and $B$ switching their apparatuses to settings not agreed upon, but coinciding with $\pm \mathbf{u}$, then one can conclude that either 
their free will is limited or that the previous decision of the sequence $\mathbf{a},\mathbf{b}$ influenced the future distribution of $\mathbf{u},\mathbf{v}$ at the entangler. 
At the same time, the latter hypothesis can be ruled out by having a third party to switch their apparatuses at the last moment.  

\section{Conclusions and Perspectives}
Quantum correlations can be obtained by local classical resources, provided the choice of setting is correlated to the  hidden variables through 
shared randomness or alleged action-at-a-distance.    
If an external observer has no access to the hidden variables,  he or she can be induced to believe that genuine quantum correlations are being observed.  
Classical resources, however, can only provide shared pseudo-randomness, so that an infinitely intelligent observer may, in principle, 
discover the underlying algorithm, even without access to the hidden variables. 
\footnote{Any pseudo-random algorithm wll repeat after a sufficiently large time, possibly several orders of magnitude the age of the universe even for a 
reasonably fast rate. However, since we are discussing a matter of principle, we may say that an infinitely patient and eternal observer, even though not necessarily intelligent, would 
eventually find out that the sequence is not actually random.}
On the other hand, if the external observer can measure the latter ones, he or she can discriminate if 
 the correlations between the hidden parameters and the settings are due to a conspiracy, a lack of free will, or an action-at-a-distance. 
 
Dropping the hypothesis of Uncorrelated Choice, however, is not the only way to reproduce the quantum mechanical correlations. 
One may relax the hypothesis of Setting Independence, instead, as in Ref.~\citeonline{Toner2003}, where the correlations are simulated 
by having one party to transmit an information $c$ to the other one. 
We have shown that this model has a natural corresponding model satisfying Setting Independence and violating Uncorrelated Choice, if one 
sees $c$ as an additional hidden variable. 

A third possibility consists in relaxing the hypothesis of Reducibility of Correlations.\cite{DiLorenzo2011g} 
I believe this last alternative leads to models as counter-intuitive as quantum mechanics, since they violate 
the classical reductionist assumption that the whole is the sum of its parts and that correlations are an expression of ignorance. 
{\section*{Acknowledgements}
This work was performed as part of the Brazilian Instituto Nacional de Ciência e
Tecnologia para a Informação Quântica (INCT--IQ) and 
it was supported by Funda\c{c}\~{a}o de Amparo \`{a} Pesquisa do 
Estado de Minas Gerais through Process No. APQ-02804-10.}

\appendix{Definitions of locality}\label{app:locdef}
It is commonly encountered the statement that the violation of Bell inequalities implies that reality is non-local. 
As far as I know, locality can have any of the following meanings
\begin{description}
\item[a.] Impossibility of instantaneous action-at-a-distance (no action-at-a-distance, NAD). 
\item[b.] Impossibility of instantaneous communication between two space-like separated parties (no-signaling, NS).
\item[c.] Impossibility for a body to have an arbitrary velocity (no faster-than-light speed, NFL).
\end{description}
These three formulations are non-equivalent, and one should specify which one is intended when using the term ``locality". 
For instance, the violation of (c) certainly implies the violation of (a) and (b), but the \emph{vice versa} is not true; 
the violation of (b) implies the violation of (a), e.g., one could send a set of instructions to a far-away robot, that would 
execute some action accordingly; I surmise that it is possible to violate (a), without violating (b), even though 
I cannot provide presently a satisfactory example. 
Hence, the chain of implications holds: $\mathrm{NAD} \implies \mathrm{NS} \implies \mathrm{NFL}$.
We shall use the stricter definition of locality, NAD. 
We notice that there is a problem in defining what is an \lq\lq{}action\rq\rq{} for a stochastic theory,  
thus we propose a suitable extension of this concept. 
First, we need to distinguish between local parameters, that Bell termed \lq\lq{}beables\rq\rq{}, and global ones. 
Local parameters are attached to points in space-time, and give rise to events when they are measured in a single shot. 
Global parameters describe a preparation procedure, and cannot be associated with the particles composing the system. 
A measurement gives a collection of events $\{e_j\}$ each corresponding to a different point of space-time. 
We say that a model is local (NAD) if the marginal probability of observing an event $e_A$ at a space-time point $A$ 
depends on global variables and only on those local parameters at the space-time point $(x_A,t_A)$.

There is a fourth definition of non-locality referring specifically to a theory predicting events $e_j$ observed by local detectors $\Sigma_j$ at space-like separated regions 
$R_j$ given that a system is specified by some parameters $\lambda$, 
\begin{description}
\setcounter{enumi}{3}
\item[d.] A theory is local if the probability of observing the events $e_j$ factorizes as 
\begin{equation}\label{eq:fact}
P(\{e\}|\lambda,\Sigma)=\prod_j P(e_j|\lambda,\Sigma_j)
\end{equation} 
\end{description}
In the special case of a bipartite entangled two-level system, we have that $\Sigma_j$ represent (pseudo)spin components, and $e_j=\pm 1$ in appropriate 
units. 
We want to show that if one takes Eq.~\eqref{eq:fact} as the definition of  locality then models which are patently local according to any of the definitions (a)-(c) are classified as non-local according to (d). 
Let us first rewrite the left hand side of Eq.~\eqref{eq:fact} by applying repeatedly Bayes's theorem
\begin{equation}\label{eq:factbayes}
P(\{e\}|\lambda,\Sigma)=\prod_j P(e_j|E_j,\lambda,\Sigma),
\end{equation} 
where $E_1=\emptyset$ is the empty set, and $E_j=\{e_1,e_2,\dots,e_{j-1}\}$. 
Eq.~\eqref{eq:factbayes} is a mathematical identity following from the rules of probability theory, and does not involve any physical assumption. 
Now, let us consider the first factor in the right hand side of Eq.~\eqref{eq:factbayes}, $P(e_1|\lambda,\Sigma)$. 
If we invoke locality (NAD) as defined above, we have the physical equality $P(e_1|\lambda,\Sigma)=P(e_1|\lambda_G,\lambda_1,\Sigma_1)$. 
The second factor, however, is 
$P(e_2|e_1,\lambda,\Sigma)$. Invoking again (NAD), we have that the marginal probability is 
$P(e_2|\lambda,\Sigma)=P(e_2|\lambda_G,\lambda_2,\Sigma_2)$. However, nothing can be said about the conditional probability 
$P(e_2|e_1,\lambda,\Sigma)$. 
Analogously, we have in general that 
\begin{equation}
P(e_j|E_j,\lambda,\Sigma)=P(e_j|E_j,\lambda,\Sigma_1,\dots,\Sigma_j).
\end{equation} 
Thus, by applying (NAD) we cannot justify Eq.~\eqref{eq:fact}. In other words, condition (d) is stronger than (NAD). 
This was pointed out by Jarrett\cite{Jarrett1984}. 
In order to obtain the equality in Eq.~\eqref{eq:fact} 
we have to make a further hypothesis, which Jarrett referred to as completeness,\cite{Jarrett1984}  
and Shimony as outcome-independence.\cite{Shimony1990}  
This hypothesis is simply that
\begin{equation}\label{eq:genoi}
P(e_j|E_j,\lambda,\Sigma_1,\dots,\Sigma_j)=P(e_j|\lambda,\Sigma_1,\dots,\Sigma_j) ,
\end{equation}
i.e., the conditional probability of observing $e_j$ given that $e_1,\dots,e_{j-1}$ were observed is identical 
to the marginal probability of observing $e_j$. 
Bell\cite{Bell1976a} tried to argue that a seemingly plausible definition of \lq\lq{}local causality\rq\rq{} 
implied Eq.~\eqref{eq:genoi}, but his proposal was criticized.\cite{Shimony1976a} 

Now, while in order to establish $P(e_j|E_j,\lambda,\Sigma_1,\dots,\Sigma_j)$ it is necessary that observers 
in $R_1,\dots,R_{j-1}$ communicate their results to the observer in $R_j$, the marginal $P(e_j|\lambda,\Sigma_1,\dots,\Sigma_j)$ 
can be determined by means of a local measurement in $R_j$, without need for communication. 
Then, assuming that Outcome Independence holds, it is now possible to invoke once again (NAD) and obtain finally 
\begin{equation}\label{eq:oiplusloc}
P(e_j|E_j,\lambda,\Sigma_1,\dots,\Sigma_j)= P(e_j|\lambda,\Sigma_j),
\end{equation}
which, upon substitution in the right hand side of Eq.~\eqref{eq:factbayes} yields Eq.~\eqref{eq:fact}. 
So far, we have basically reformulated the conclusions of Ref.~\citeonline{Jarrett1984}, that, however, is not widely known to physicists. 

Let us establish sufficient conditions for the validity of Outcome Independence as formulated in Eq.~\eqref{eq:oi}. 
Determinism is a sufficient condition: if the knowledge of $\lambda$ determines the outcome of the measurements $\Sigma_j$, all other information is redundant.
Another sufficient hypothesis is that of probabilistic determinism, which is not an oxymoron, but 
means that knowledge of the $\lambda$ determines the probability of the outcome, not the outcome itself. 
Perhaps this concept coincides with what Jarrett\cite{Jarrett1984} calls ``completeness'', which we avoid since it is a term laden 
with subjective meanings. 
Another sufficient condition is separability:\cite{Howard1989} the parameters $\lambda=\bigcup_j \lambda_j$, where $\lambda_j$ 
are parameters attached to the particle number $j$ and represent prepossessed values. 
In other words, there are no global parameters. 
Then we have that the probability of any outcome $e_j$
can be determined by knowledge of the $\lambda_j$ alone, i.e., 
$P(e_j|E_j,\lambda,\Sigma_1,\dots,\Sigma_j)=P(e_j|\lambda_j,\Sigma_j)$, which is a special form of Outcome Independence. 
Bell, in particular, assumed\cite{Bell1976a} that all hidden parameters where of the local type (the \lq\lq{}beables\rq\rq{}), and that global parameters were 
an expression of ignorance of \lq\lq{}beables\rq\rq{}, and could thus be eliminated. 
Finally, it can be proved\cite{Hall2011} that if a model satisfies factorability then there is a natural extension of this model which is 
deterministic. 

In order to show the absurdity of identifiying Eq.~\eqref{eq:fact} with locality, we shall provide now a simple model which is manifestly local, but  
would be classified as non-local according to (d). 
Let us consider the following experiment: 
a dealer prepares two decks of cards which can be a King ($K$) or a 
Queen ($Q$), and Black ($B$) or Red ($R$); each deck is 
subdivided into pairs, such that each pair is formed by a King and 
a Queen, and a Black and a Red card. 
In the first deck ($D_1$), 30\% of the pairs are $(KR,QB)$, and 70\% $(KB,QR)$. 
In the second deck ($D_2$), 70\% of the pairs are $(KR,QB)$, and 30\% $(KB,QR)$. 
The dealer chooses a deck at random, with equal probability, then extracts a pair out of the deck, 
and handles one card each to two observers, one, $A$, sitting to his left and the other, $B$, to his right. 
In this model, the hidden variable $\lambda$ is the deck which has been chosen. 
Once the pair is extracted from the deck, $\lambda$ is not localized on either card, but is a global parameter, which cannot 
be reconstructed by observers $A$ and $B$ by making local observations (even if they compare their results). 
The only way to determine $\lambda$ would be to check which deck was chosen at the location of the preparation. 
Now let us consider the joint probability that $A$ will receive a King and $B$ will receive a black card
given that deck 1 was chosen: 
\begin{equation}
P(K,B|\lambda=D_1)= \frac{3}{20} ,
\end{equation}
since a pair with a red King and a black Queen will be extracted with probability $3/10$, and the black Queen is received by $R$ half of the times.  
On the other hand, if factorability holds, we should have 
\begin{equation}
P(K,B|D_1)= P(K|D_1)P(B|D_1)=\frac{1}{4}. 
\end{equation}
Thus, according to definition (d), the model we have presented is non-local. 

Notice that, elaborating further our example (or with any example which we can make out with our 
classical imagination), it is not possible to violate 
Bell inequality. The reason lies in the fact that in any classical 
case the hypothesis of counter-factual definiteness is always satisfied. 
The peculiarity of reality revealed by the experimental violation of Bell inequalities consists in the fact that no 
such complete characterization of a system exists in principle. 

\begin{figure}
\centering
\includegraphics[width=4in]{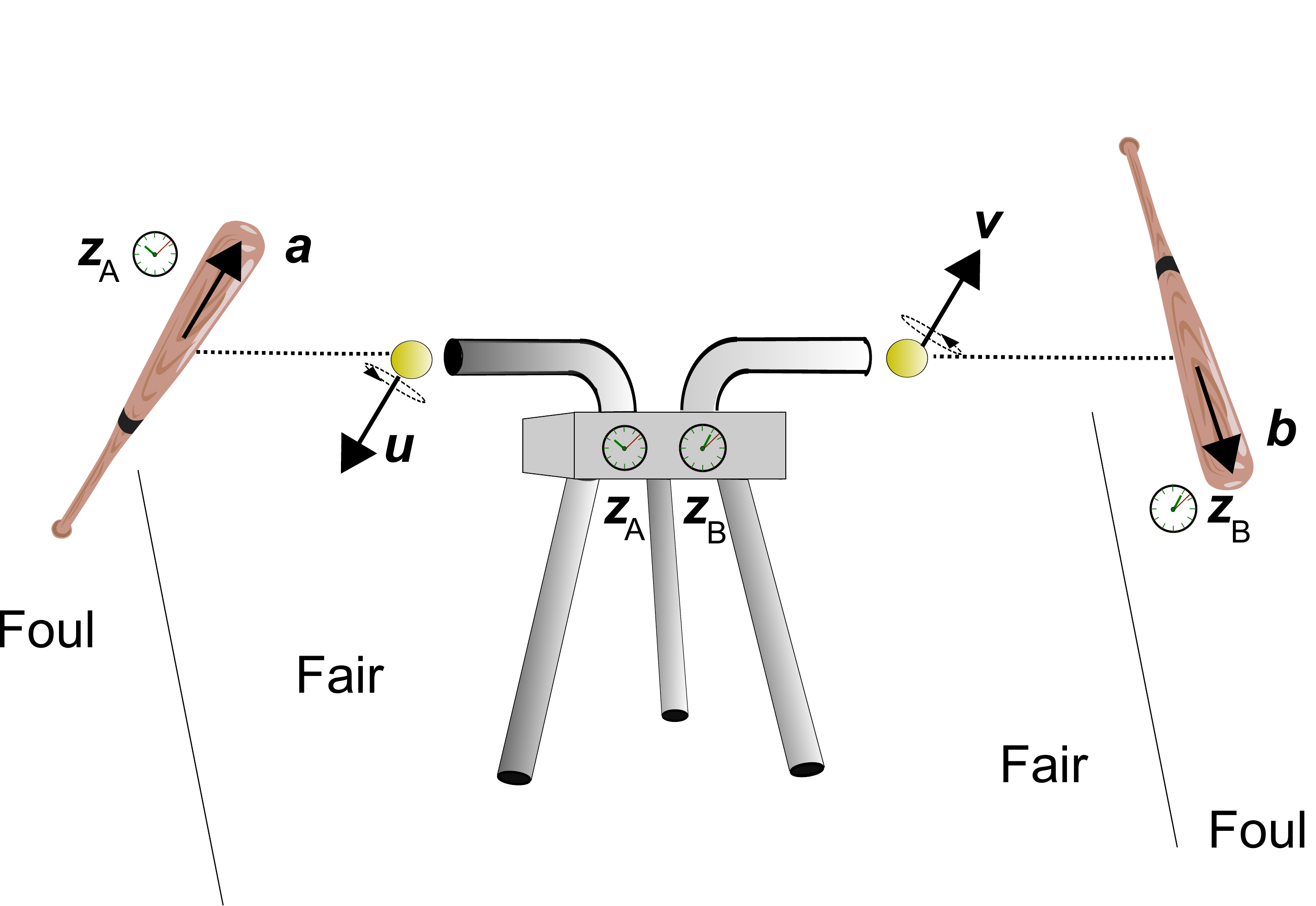}
\caption{\label{fig:setup} Scheme of the setup.}
\end{figure}

\begin{figure}
\centering
\includegraphics[width=4in]{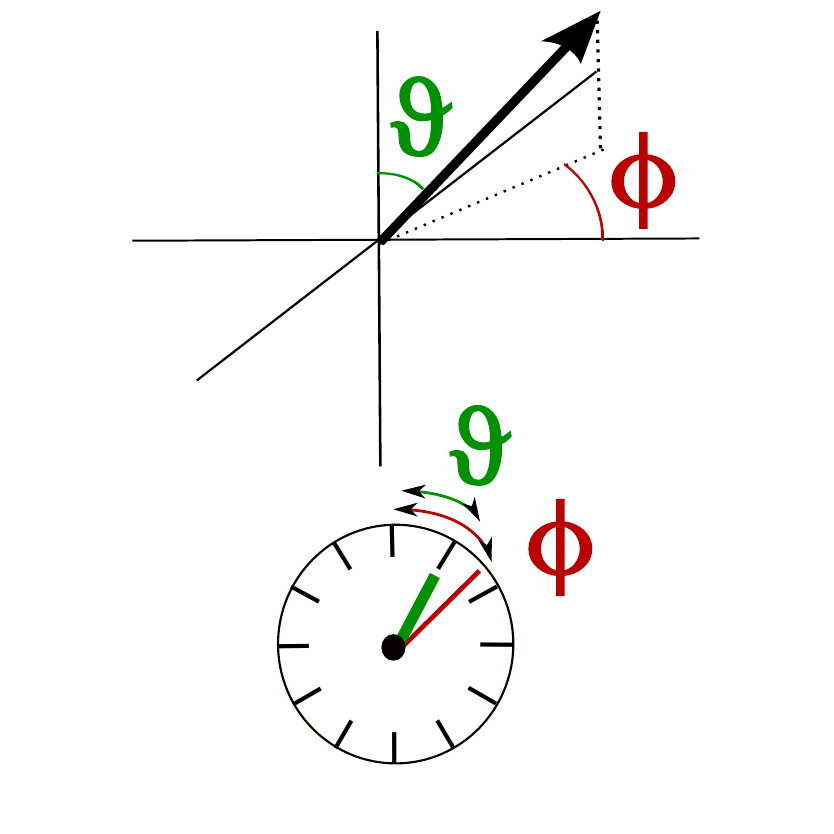}
\caption{\label{fig:watch} A unit vector is associated with the hands of a watch.}
\end{figure}

\appendix{Realization of the model of Section \ref{sec:model} through entangler-detectors shared randomness}\label{app:real}
Here we present a further realization of the model of Section \ref{sec:model}, 
without having any shared randomness between $A$ and $B$, but only between each station and the entangler. 
Let us consider the following setup depicted in Fig. \ref{fig:setup}: 
there is a sophisticated baseball pitching machine able to pitch two balls spinning with angular velocities 
 $\boldsymbol{\omega}$  and $-\boldsymbol{\omega}$ (with $\omega$ fixed)
in opposite directions. The balls are approximatedly a rigid body and have spherical symmetry. 
They are pitched with fixed center-of-mass velocities in a vacuum (so that the spin does not 
influence the center-of-mass trajectory through the Magnus effect) 
such that their centers of mass follow given trajectories each ending 
in a bat controlled by an independent system. 
Each bat is carefully crafted to be a solid of revolution, its center of mass is being held fixed, 
and a machine varies the orientation $\mathbf{n}$ of the bat's symmetry axis.    
It is empirically found that a ball pitched with $\boldsymbol{\omega}=\omega \mathbf{u}$, 
after hitting a bat whose axis is oriented along $\mathbf{n}$, will fall in the foul ground with 
a frequency $(1-\mathbf{n}\cdot\mathbf{u})/2$, and in the fair ground with the complementary 
frequency $(1+\mathbf{n}\cdot\mathbf{u})/2$. 
Due to our politically incorrect bias against negative numbers, 
we shall associate the value $\sigma=-1$ to the event of the ball being batted in the foul ground, 
and $\sigma=+1$ to the alternative event. 
The pitching machine varies the spin of the two balls according to the following algorithm: 
it flips a fair coin, and according to the result, heads $H=A$ or tails $T=B$, 
it consults either of two internal watches, $W_A$ and $W_B$. 
Each watch is built so that the small hand has a period $\tau_{j,s}$ and the large hand 
$\tau_{j,l}$, with $j\in\{A,B\}$. Unlike what happens in ordinary watches, 
any two of the four periods are mutually incommensurable.
The positions of the hands are used to determine a unit vector $\mathbf{z}_j$. 
This may happen in a straightforward fashion, as depicted in Fig. \ref{fig:watch} 
or by using a Montecarlo algorithm, 
so that an external observer could not predict what spin is chosen, even knowing 
the periods of the watches. 
The machine will then pitch a ball spinning along $\mathbf{u}=\mathbf{z}_j$ to the left, and one 
spinning along $\mathbf{v}=-\mathbf{z}_j$ to the right, or \emph{viceversa}, according to another 
coin flip, with outcome $d\in\{H,T\}=\{1,-1\}$. 

The batting machine to the right possesses a watch $W'_H$, and the one to the left a watch 
$W'_T$, which have the same periods as the watches of the pitching machine, 
and are synchronized with them. 
Each batting machine is using the same algorithm as the pitching machine to determine 
the value of $\mathbf{z}_j$, but it is subtracting 
from the input values $t_{j,h}$ the number $\Delta t \ (\mathrm{mod}\ \tau_{j,h})$, 
where $\Delta t$ is the time-of-flight of the ball.
The batting machines will orient the bats along $\mathbf{a}=\mathbf{z}_A$ and $\mathbf{b}=\mathbf{z}_B$, respectively.  
We remind that the probability of the outcome $\sigma$
for a ball spinning in the direction $\mathbf{u}$ and hitting a bat oriented along $\mathbf{n}$ 
is 
\begin{equation}
M_{\sigma}(\mathbf{u},\mathbf{n})=\frac{1}{2}\left[1+\sigma\mathbf{n}\cdot\mathbf{u}\right] . 
\end{equation}
It is easy to check that the expected joint frequency of observing the outcomes 
$\sigma,\tau$ to the left and to the right when the respective bats have 
orientations $\mathbf{a},\mathbf{b}$ is 
\begin{align}
\nonumber
P_{\sigma,\tau}(\mathbf{a},\mathbf{b})&
=\int\! d\mathbf{u}d\mathbf{v}\ \mu(\mathbf{u},\mathbf{v}|\mathbf{a},\mathbf{b})
M_{\sigma}(\mathbf{u},\mathbf{a})
M_{\tau}(\mathbf{v},\mathbf{b})\\
&=
\frac{1}{4}\left[1-\sigma\tau\mathbf{a}\cdot\mathbf{b}\right] ,
\end{align}
with the density 
\begin{equation}\label{eq:dilorconddistr}
\mu(\mathbf{u},\mathbf{v},\mathbf{z}_A,\mathbf{z}_B|\mathbf{a},\mathbf{b})=
\frac{1}{4}\delta(\mathbf{u}+\mathbf{v})
\delta(\mathbf{z}_A-\mathbf{a})
\delta(\mathbf{z}_B-\mathbf{b})
\sum_{j,d}\delta(\mathbf{u}-d\,\mathbf{z}_j) \\
\end{equation}

We notice that the batting machines are devoid of ``free will'', since the directions $\mathbf{a},\mathbf{b}$ 
are determined, while the pitching machine has two binary choices, first in choosing either watch, 
then in choosing which ball to pitch towards, e.g., left. 

It is immediate to realize that the model of Ref.~\citeonline{Hall2010} 
can be reproduced by changing the algorithm used by the pitching machine. 
Now the pitching machine will use both watches $W_H,W_T$ 
to determine two unit vectors $\mathbf{a},\mathbf{b}$. 
The pitcher is in possession of a third watch $W_0$, whose hands determine a unit vector $\mathbf{u}$. 
The third watch, however, is not ticking regularly, but it is coupled to the watches $W_R$ and $W_L$ in such 
a way that the hands correspond to the vector $\mathbf{u}$, for given $\mathbf{a},\mathbf{b}$, with 
the frequency 
\begin{equation}\label{eq:halldens3}
\Pi(\mathbf{u}|\mathbf{a},\mathbf{b})= 
\frac{1-f(\mathbf{u},-\mathbf{u}, \mathbf{a},\mathbf{b})}{8\arccos{f(\mathbf{u},-\mathbf{u}, \mathbf{a},\mathbf{b})}}. 
\end{equation}
The pitcher will then proceed to pitch two balls, the first, spinning about $\mathbf{u}$, to the left, and the second, 
spinning about $-\mathbf{u}$, to the right. 
The other parameters of the procedure $\omega,v$ are readjusted such that a ball spinning around 
$\mathbf{u}$ hitting a bat oriented along $\mathbf{n}$ will give the outcome $\sigma$ with 
probability 
\begin{equation}\label{eq:hallcondprob}
P_\sigma(\mathbf{u},\mathbf{n})=\delta_{\sigma,\mathrm{sgn}(\mathbf{u}\cdot\mathbf{n})}.
\end{equation}
The batters, on the other hand, keep using the former algorithm in order to determine $\mathbf{a},\mathbf{b}$.

\section*{References}

\end{document}